\documentclass[useAMS,usenatbib,usegraphicx,referee]{mn2e}
\newcommand{\rg}{r_{\rm g}}
\newcommand{\taur}{\tau_{\rm r}}
\newcommand{\mybf}{}

\usepackage{amssymb}

\begin{document}

\title[P-Cygni profiles from outflows near compact objects.]
{Profiles of spectral lines from failed and decelerated winds from neutron stars and black holes.
}
\author[A. V. Dorodnitsyn ]{A. V. Dorodnitsyn$^{1}$\thanks{E-mail:
dora@milkyway.gsfc.nasa.gov}\\
$^{1}$Laboratory for High Energy Astrophysics, NASA Goddard Space Flight Center, Code 662, Greenbelt, MD, 20771, USA}

\date{}

\pagerange{\pageref{firstpage}--\pageref{lastpage}} \pubyear{2002}
\maketitle

\label{firstpage}
\begin{abstract}
We calculate profiles of spectral lines from an extended
outflow from the compact object
(a black hole, or a neutron star).
We assume that the bulk velocity of the flow increases during
a short phase of acceleration and then rapidly decreases forming a failed wind. 
We also study the wind which is only decelerating.
We show
that depending on the relative strength of the gravitational redshifting, line profiles from such winds
may be of several types: distorted P-Cygni (emission and blueshifted absorption);
W-shaped (absorption-emission-absorption); and inverted P-Cygni (emission - {\it redshifted} absorption).
The latter case is expected from accretion flows where the velocity is directed inward, however we show that inverted
P-Cygni profile can be produced by the failed wind,
provided the line is formed within several tens of 
Schwarzschild radii from the compact object. 
\end{abstract}


\section{Introduction}

Relativistically broadened, double horn profiles from fluorescent Fe K$\alpha$ lines is a widely accepted
evidence for the presence of the cold accretion disk in Seyfert 1 active galactic nucleus (AGN), 
and the assumption that such disk is illuminated by 
X-rays generated by the corona (see, e.g. \citet{Fabian2000}). However, the geometrically broad line forming region spans
a large range of orbital velocities which translates to a broad region in frequency space, additionally skewing the line by relativistic effects and making
detailed diagnostics of the accreting plasma difficult.

Spectral resolution of grating spectrographs of the X-ray telescopes {\it Chandra} and XMM-{\it Newton} below $\sim 10$ keV allows
observations revealing complicated structure of the $\rm K\alpha$ line, in many Type 1 AGNs e.g. \citep{Reeves01}; 
and for the detection of the narrow
core, see also \cite{Yaqoob2003}. 
Observations suggest that distant parts of the accretion disk as well as the reflection from the obscuring torus are involved.
Multidimensional simulations of spectra of warm absorber flows by \cite{DorodnitsynKallman09} also show the importance 
of distant AGN winds for the formation of the complex structure of Fe K$\alpha$ line.

It is widely accepted that strong gravitational field of the compact object, such as black hole (BH), or neutron star (NS) can
imprint itself into the radiation which comes from regions within several tens of Schwarzschild radii, $r_{g}$.
Narrow emission and absorption lines from black holes and neutron stars delivers an optimal opportunity to deduce their masses and 
radii (in case of NS).

Narrow lines formed near a BH carry direct information about the dynamics, ionization stage and covering fraction of the absorbing gas and thus about the mass budget of the inflow/outflow. The latter is important for understanding of the fraction of the AGN - host galaxy feedback which comes directly from regions close the BH. 
Gravitational redshifting is entangled with Doppler shifts, and altogether they encode information about the mass of the BH.

Observed narrow absorption features from several Quasars and Seyfert galaxies possibly reveal the importance of 
gravitational redshifting and Doppler blue-, and redshifting due to the bulk motion of the moving gas.
Such evidence was found in several cases including: Seyfert 1 galaxy NGC 3516 where it is suggested that observed absorption features 
are from the gravitationally 
redshifted resonant line scattering within the accretion flow \citep{Nandra99};
the gravitationally redshifted (10 - 20 $r_g$ from the BH) resonance absorption line from $\rm Fe{\,XXV}$
or $\rm Fe{\,XXVI}$  in the quasar E1821+643 \citep{Yaqoob};
a combination of the gravitational and Doppler shifting of lines from ionized Iron is suggested 
to explain absorption features from the quasar PG 1211+143 by
\citet{Reeves}; also see \citet{Matt} for possible detection of the strongly redshifted, transient absorption feature from ionized Fe in
the quasar Q0056-363. 

\cite{Dadina05} reported on the redshifted absorption, and emission features observed at $\sim 6$ keV, from Seyfert 1 galaxy Mrk 509. Two of the absorption features are 
separated by the emission line, and the overall emission structure is attributed to H-, or He-like Iron. This W-shaped structure is found to be sporadic with variability as short as $\sim~20$ ks. Infall and outflow, $\sim\pm \, (0.1 - 0.2)\,c$ of matter are suggested, together with the gravitational redshifting to explain these features.
Transient nature of such features is consistent with the latter non-detection of the redshifted absorption and the detection of only blueshifted absorption lines
\citep{Cappi09}.

The detection of the redshifted absorption lines form accreting neutron stars which exhibit thermonuclear X-ray bursts are ideally suited for deducing 
the NS mass and putting constrains on its equation of state. So far, these detection are rare. \cite{Cottam02} identified several absorption features in the burst spectra of the neutron star EXO0748-676. Lines from H-like Iron for the early phases of the burst, and He-like Iron for the late phases were adopted to explain the spectra, and to obtain the redshift of ${\it Z}=0.35$.  Other studies, including LTE and non-LTE neutron star atmosphere modeling suggested
$n=2-3$ transition of $\rm Fe{\,XXIV}$, and the redshift ${\it Z}=0.24$ \citep{Rauch08}.
These lines are expected to be intrinsically transient which additionally lowers
the statistical significance of such detections.

Usually, the existing models of gravitationally redshifted lines do not include 
influence of the transfer effects in the extended
moving envelope, and if they do, simple transmission models are adopted.
Large intensities of emission components, and simultaneous presence of emission and absorption from the same line transition
suggest the importance of the coupling between dynamics of the wind and gravitational redshifting.

Beginning from works of \cite{Beals} it was understood
that P-Cygni profiles provide a unique evidence for the rapidly moving wind. It was also understood that radiation from a huge volume, occupied 
by such wind naturally explains bright emission features in the spectra, and the observed broadening is due to high velocity dispersion of such wind
(${\rm few} \times 10^{3} \,{\rm km\,s^{-1}}$ in the case of a normal star). The most remarkable about this profile is, of course, the relative blueshifting of the absorption trough relative to the emission line. 
As a P-Cygni profile bears an imprint of the bulk motion of the plasma, it is an important tool of the diagnostics of the dynamics of stellar winds.

In the paper by \cite{Dorodnitsyn09} (hereafter Paper I)
a modified Sobolev approximation was built which incorporate gravitational redshifting, 
and spectral line profiles were calculated from a stellar-type and explosion-type winds, accelerated in the strong gravitational field of the compact object.

An important parameter from Paper I measures the relative importance of gravity at the wind base $g_{0}=R_{\rm c}/r_{g}$, 
where $R_{\rm c}$ is the radius of the wind launching point.
Two types of the velocity laws were adopted: i) ''stellar'' type, i.e. $v(r)\sim(1-R_{c}/r)^{m}$, and ii) Homologous expansion $v(r)\sim r$ (Hubble law), adopted for the description of outbursts.
Important also found the terminal velocity, $V^{\infty}$, which spans a range of 0.01- 0.3 c and the distribution of the opacity. 

It was found that gradually accelerated winds which are launched at $R_{c}\lesssim 20-30\,r_{g}$ 
produce the following types of profiles:
i) distorted P-Cygni, i.e. characterized by emission feature which is redshifted relative to an absorption trough 
ii) saw-tooth, i.e. the narrow redshifted absorption line superimposed on a broader redshifted emission line, 
combined additionally with a blueshifted absorption line 
iii) W-shaped, i.e. absorption-emission-absorption profiles
{\mybf (by ''red-shifted'' or 'blue-shifted'' we necessarily mean only the relative blue-, red-shift of one spectral feature with respect to the other,
which may, or may not represent the actual shift of frequency of such feature with respect to the rest frequency of the line)}.

In this paper we calculate profiles of lines from a wind which is either everywhere decelerated, i.e. described by the $v\sim 1/r^{m}$ law 
or failed, i.e. its velocity rises steeply during the short acceleration phase, and then decreases, asymptotically approaching $v\sim 1/r^{m}$ law.

We will show that line profiles from failed winds are significantly different from those considered in Paper I. 
Additionally to three types described above there is a forth type:
the inverted P-Cygni profile (absorption-emission).
Notice, that the inverted P-Cygni profile is expected from spherically-symmetric accretion flow, and to observe it from the wind is counter-intuitive. 
We will show that 
strong gravitational redshifting together with the ''failed'' character of the wind, and also
provided the deceleration phase 
is fast (i.e. $m\gtrsim 3$) are 
necessary conditions for the appearance of this feature.
Failed wind profiles can also be W-shaped or saw-tooth shaped.

The plan of this paper is as follows: in Section 2 we review basic assumptions of Paper I about the wind geometry and gravitational field; 
in Section 3 we describe our methods of calculation of line profiles i.e. the Sobolev optical depth of the line, source functions, calculate mean radiation field etc.;
in Section 4 we describe velocity laws, derive the bridging formula for the failed wind and make assumptions about the distribution of the opacity;
in Section 5 we calculate equal frequency surfaces, which are of paramount importance to understand line profiles which we calculate in Section 6;
in Section 7 and 8 we discuss major results and conclude.

\section{Assumptions and approximations}

We assume that photons from an external source of continuum radiation are interacting with the wind via resonance line scattering; 
both the wind and the radiation source are assumed to be spherically symmetric; 
the wind is launched from the photosphere
of the radius $R_{\rm c}$; 
there is a prescribed velocity profile, $v(r)$ and a prescribed distribution of the line opacity in the wind. 
The only difference in the assumptions with those of Paper I is that here we consider decelerating winds, instead of winds which are gradually increasing their velocity. Profiles from decelerated winds from normal stars were calculated in works of different authors (e.g. \citet{KuanKuhi, Marti77}).
In the following we briefly summarize the formalism developed in Paper I.

We assume that
after leaving the photosphere a photon is traveling in the wind without interaction with the matter except for certain points (resonances) where it's frequency in the co-moving frame, $\tilde\nu$ appears to be within the Doppler thermal width of the spectral line, $\Delta\nu_D=\nu_{0}\, v_{th}/c$, where $v_{th}$ is the thermal velocity, and $\nu_{0}$ is the frequency at the line center. To allow for the resonance region to be located as close as several Schwarzschild radii, $r_g=2GM/c^{2}$ from the compact  object  we take into account both Doppler and gravitational shifting of the photon's frequency. 
{\mybf The Lorentz transformations from
a local Lorentzian frame (a photon frequency $\nu_{\rm loc}$) which is at rest at a given point $s_0$
to the co-moving frame give:
${\tilde \nu}=\gamma\nu_{\rm loc}(1-\mu\beta)$, where $\beta\equiv
v/c$, $\gamma\equiv(1-\beta^2)^{-1/2}$.
A photon that was emitted  at the point $s_0$ after traveling 
to some other point $s$ is gravitationally red-, blueshifted and in 
the co-moving frame has the following frequency:
${\tilde \nu}(s)=
\gamma\nu_{\rm loc}(s_0)\sqrt{g_{00}(s_0)/g_{00}(s)}\,\left(1-\mu(s)\beta(s)\right)$.
}

Gravitational field is described in a ''weak field limit'', i.e. by means of the  gravitational potential, $\phi$, {and also
implying $\sqrt{g_{00}}\simeq 1+\phi/c^2$}.
To allow for the terminal velocity, $V^\infty$ to be of the order of the escape velocity $V_{\rm esc}$ at the base of the wind (which can be a large fraction of the speed of light, $c$), we must retain all terms of the order of $v^2/c^2$ and $\phi/c^2$.

The photon emitted at the point $s$, in a direction ${\bf n}$, having the frequency $\tilde \nu$ will be registered by the observer at infinity at the frequency $\nu^\infty$:

\begin{equation}\label{nu_infty_vs_nu_com}
\tilde\nu(s)=\nu^\infty\left(1-\mu(s)\beta(s)-\frac{\phi(s)}{c^2}+
\frac{\beta(s)^2}{2}\right)\mbox{,}
\end{equation}
where $\mu={\bf n}\cdot {\bf v} = \cos\theta$, and $\phi$ is the gravitational potential.

It is convenient to adopt dimensionless units: the non-dimensional frequency, $y= (\nu-\nu_0)/\Delta\nu_{\rm D}$, the non-dimensional radius,
$x=r/R_{\rm c}$, 
the non-dimensional velocity, $u=v/v_{\rm th}$, and the non-dimensional gravitational potential $\Phi=\phi/(c\, v_{\rm th})=\zeta\,(\phi/c^{2})$, where $\zeta= c/v_{\rm th}$.
In the narrow line limit, the emitted frequency is $\nu_{0}$ and thus

\begin{equation}\label{yinfty_definition}
y^{\infty}= (\nu^{\infty}-\nu_0)/\Delta\nu_{\rm D}\mbox{.}
\end{equation}
In such non-dimensional units, the equation (\ref{nu_infty_vs_nu_com}) can be cast in the form:

\begin{equation}\label{y_infty_main_equation}
y^\infty=\mu\,u(x)+{\Phi}(x)-u(x)^2/(2\zeta)\mbox{.}
\end{equation}
Equation (\ref{y_infty_main_equation}) determines the position of the resonant point.
It is straightforward to rewrite the above equations in terms of the redshift,
${\it Z}=(\tilde\nu-\nu^{\infty})/\nu^{\infty}$:

\begin{equation}\label{express_yinfty_Z}
y^\infty=-\zeta\frac{\it Z}{1+{\it Z}}\mbox{,}
\end{equation}
and from equation (\ref{y_infty_main_equation}), we obtain:

\begin{equation}\label{Z_main_equation}
{\it Z}= - \frac{\mu\,u+\Phi - u^{2}/(2\zeta) }
  {\zeta + \mu\,u+\Phi-u^{2} / (2 \zeta) }\mbox{.}
\end{equation}

To describe gravitational field
we adopt two types of potentials: the Newtonian potential, $\phi(r)=-\frac{GM}{r}$, and the pseudo-Newtonian potential of Paczynski-Wiita
(PW) \citep{PaczWiita}: 

\begin{equation}\label{PaczWiita_potential1}
{\displaystyle \phi(r)=\frac{GM}{r_g-r}}\mbox{.}
\end{equation}
The latter mimics important features of exact general
relativistic solutions for particle trajectories near a Schwarzschild
black hole. For example, it reproduces the positions of
both the last stable circular orbit, located at $3\rg$ and the
marginally bound circular orbit at $2\rg$.
The non-dimensional form of the PW potential, (\ref{PaczWiita_potential1}) reads: 
${\displaystyle {\Phi(x)}=\zeta/ \left( 2(1-x
g_0)\right)}$.
A non-dimensional parameter,
$g_0= R_{\rm c}/\rg$ determines the relative
importance of the gravitational redshifting (i.e. by equating
$g_0\to\infty$ one completely neglects the influence of the
gravitational field on the energy of a photon). 

We adopt a $(p,z)$ coordinate system, such that $x=\sqrt{p^{2}+z^{2}}$, where $p$ is the impact parameter and the observer is located at $z=\infty$.
In the non-dimensional variables, equation (\ref{y_infty_main_equation}) takes the form:

\begin{equation}\label{y_infty_main_eq_nondim}
y^\infty=\frac{z_0}{\sqrt{z_0^2+p^2}}
u\left(\sqrt{z_0^2+p^2}\right)+\Phi\left(\sqrt{z_0^2+p^2}\right)
-\frac{1}{2\zeta} u^2\left(\sqrt{z_0^2+p^2}\right)
\mbox{.}
\end{equation}
The solution of the 
equation (\ref{y_infty_main_eq_nondim})
for different values of $p$ determines the locus and shape of the equal frequency surface (hereafter EFS). 
Understanding the shape, and topology of EFS is very helpful in order
to understand and interpret profiles of spectral lines from the moving medium.
Notice, that equation, (\ref{y_infty_main_eq_nondim}) does not reference to any  particular form of the velocity law, $v(r)$. 
In the absence of gravitational redshifting, the first term on the right hand side of
(\ref{y_infty_main_eq_nondim}) determines the locus of the surfaces of equal line-of-sight velocities. The second and third terms, stand for the gravitational redshifting and transverse Doppler effect (also redshifting). Equation (\ref{y_infty_main_eq_nondim}) was derived in Paper I, where its properties are investigated for the case of the outwardly accelerated wind. Here we study this equation from the perspective of decelerated winds.

\section{Calculation of line profiles: methods}
{\mybf To include gravitational redshifting into the frame of the Sobolev approximation, requires 
to take into account terms of the order ${\cal O}(v^{2}/c^{2})$.
Sobolev optical depth enters Castor's \citep{Castor70} escape probabilities, thus not only making them dependent  on $d\phi/d{\bf s}$, but
also demanding to include ${\cal O}(v^{2}/c^{2})$ terms into their derivation. A Sobolev approximation e.g. \citep{Sob, RybickiHummer78} which takes into account all of the above assumptions, 
was developed in Paper I, and including decelerating wind into consideration leaves this formalism intact. 
}
The Sobolev length, $L_{\rm sob}$ is defined as a characteristic length over which the photon, propagating in the direction ${\bf s}$ stays in resonance with a certain line transition:
$\displaystyle L_{\rm sob} \simeq v_{\rm th} / \left| {dv}/{d{\bf s}} +(1/c)\, {d\phi}/{d{\bf s}}\right|$. The optical thickness  in the direction
${\bf s}$ is 
\begin{equation}\label{tau_sob_def}
\tau_{\rm sob} = \chi^l_{0,\rm com}\times L_{\rm sob}\mbox{.}
\end{equation}

\noindent
The Sobolev optical depth in a line is found from the relation:

\begin{eqnarray}\label{tau_Line_Sob}
\tau_{\rm l}&=& \chi^l_{0,\rm com}\frac{r}{\beta}\left(1+\beta^2(1+\mu^2)-2\mu\beta-\frac{\phi}{c^2}\right)\\
&   &\left[
1-\mu^2\left(1-\frac{d\ln \beta}{d\ln r}\right)\right.
+\left.\mu \left( \frac{1}{\beta\,c^2} \frac{d\phi}{d \ln r} -\beta\frac{d\ln \beta}{d\ln r}
\right) \right]^{-1}
\mbox{,}\nonumber
\end{eqnarray}
where $\chi^l_{0,\rm com}$ is the opacity at the line frequency in the co-moving frame:

\begin{equation}\label{kappa_com}
  \chi^l_{0,\rm com}=\frac{\pi e^2}{m c} (gf)
  \frac{N_l/g_l-N_u/g_u}{\Delta\nu}\mbox{,}
\end{equation}
where $N_u$, $N_l$ and $g_u$, $g_l$ are populations and
statistical weights of the corresponding levels of the transition, and $f$ is the oscillator strength of the transition. Retaining terms of the order ${\cal O}(v^{2}/c^{2})$, we recast equation (\ref{tau_Line_Sob}) in the form:

\begin{eqnarray}\label{tau_Line_Sob_Taur}
\tau_{\rm l}&=& \tau_{\rm r} \times\left(\frac{d\ln\beta}{d\ln r}\left(1+\beta(1-2\mu)\right)+\frac{1}{c^2\beta}\,\frac{d\phi}{d\ln r}\right)\\
&   &\times\left[
1-\mu^2\left(1-\frac{d\ln \beta}{d\ln r}\right)\right.
+\left.\mu \left( \frac{1}{\beta\,c^2} \frac{d\phi}{d \ln r} -\beta\frac{d\ln \beta}{d\ln r}
\right) \right]^{-1}
\mbox{,}\nonumber
\end{eqnarray}
where $\tau_{\rm r}=\tau_{\rm l}(\mu=1)$ is the optical depth in the radial direction. {\mybf The Sobolev approximation consists of two parts: i) calculation of the optical depth of a resonant layer (i.e. from {(\ref{tau_Line_Sob}))}, and 
ii) calculation of the source function $S_{\nu}$. Considering the scattering within a line and adopting complete frequency redistribution, and isotropic scattering matrix, the source function can be approximated as 
(i.e. \cite{Hummer69}):

\begin{equation}\label{SourceFunctionfromScattering}
S_{\nu}\simeq (1-\epsilon)\,J_{\nu} + \epsilon B_{\nu}\mbox{,} 
\end{equation}
where $J_{\nu}$ is the mean intensity, and $B_{\nu}$ is the Planck function.
$\epsilon$ is the ratio of collisional and de-exitation rates.} 

In order to calculate the source function and then, finally, the emergent intensity, we adopt a method of escape probabilities of \cite{Castor70}. 
In this approach, the mean intensity $J_{\nu}$ is found {\mybf from averaging of the formal solution of the radiation transfer equation over all solid angles;  the result is}

\begin{equation}\label{MeanIntensity_EscProb}
J_{\nu}=S_{\nu}(1-P_{\rm esc}) + J_{\rm dist}(P_{\rm pen})\mbox{,}
\end{equation}
where $P_{\rm esc}$ is the probability for a photon to escape from the resonant region, and $J_{\rm dist}$ is the contribution to the mean radiation field delivered by the distant core,
and $P_{\rm pen}$ is the probability for a photon, which is issued by the core, to 
penetrate to the given point. {Combining (\ref{SourceFunctionfromScattering}) and (\ref{MeanIntensity_EscProb}) in a single linear equation and
solving for $S_{\nu}$, we obtain: $S_{\nu}\sim J_{\rm dist}(P_{\rm pen})/P_{\rm esc}$ (notice, that we assumed $\epsilon=0$).}
The probability for a photon to escape in a direction $\mu$ reads:

\begin{equation}\label{beta_esc}
P_{\rm esc}=\langle P^\mu_{\rm loc} \left(1+\beta^2(3\mu^2-1)+2\mu\beta\right) \rangle
\mbox{,}
\end{equation}
where $\langle\rangle$ denotes angular averaging: ${\displaystyle \langle f\rangle_{\Omega}= \frac{1}{4\pi}\int_{\Omega}\,f(\Omega')\,d\Omega'}$, and
$P^\mu_{\rm loc}$ is the non-relativistic expression for the directional escape probability:

\begin{equation}\label{beta_mu}
P^\mu_{\rm loc}=\frac{1-\exp(-\tau_{\rm l}(\mu,s))}{\tau_{\rm l}(s,\mu)}
\mbox{,}
\end{equation}
where $\tau_{\rm l}$ is obtained from (\ref{tau_Line_Sob}). 
We assume that the core emits continuum black-body radiation, with no limb darkening, and that in the vicinity of the line the continuum may be approximated as constant. 
Integrating $P^\mu_{\rm loc}$ over the maximum angle subtended by the core:
$\displaystyle \theta_c=\arccos{\mu_c}= \arccos{\sqrt{1-x^{-2}  }}$, the probability
for a photon to penetrate to a given point reads:

\begin{equation}\label{beta_pen}
P_{\rm pen}=\langle P^\mu_{\rm loc}\rangle_{4\pi W}\mbox{,}
\end{equation}
where $W$ is a dilution factor:

\begin{equation}\label{dilution_factor}
W=\frac{1}{2}\left(1-\mu_c\right)\mbox{.}
\end{equation}
The distant contribution,
$J_{\rm dist}$ is calculated from

\begin{equation}\label{J_distant}
J_{\rm dist}=I_{\rm c}\, P_{\rm pen}\mbox{,}
\end{equation}
and for the source function, $S_{\nu}$ we have
\begin{equation}\label{SourceFunction}
S_{\nu} = \frac{J_{\nu}}{P_{\rm esc}} \left(1+\beta^2(3\mu^2-1)+2\mu\beta\right)\mbox{.}
\end{equation}

\noindent
{\mybf In the presence of non-monotonic  velocity law, for example when the wind first accelerates and then decelerates, multiple resonances may occur. This are best studied adopting 
a concept of equal frequency surfaces (see next section).} 
At the $i^{th}$ resonance, the radiation field suffers a change: some photons from directions other than that of the ray, are scattered into the direction of the observer (local contribution); the other contribution is due to the incident intensity which is attenuated at this point:

\begin{equation}\label{Intens_jump_onEFS}
{I}_{\rm res}={I}_{\rm inc} e^{-\tau_{{\rm l},i}}+
{S_{i}}(1-e^{-\tau_{{\rm l},i}})\mbox{,}
\end{equation}
where $S_{i}$ is the source function at the $i^{th}$ resonance.
The source function $S_{i}$ is anisotropic in the direction of motion and was calculated in the relativistic case
in the paper by \cite{HutchSurd95}. {\mybf  Notice, that these authors calculated the full relativistic source function, and our calculations take into account the contribution from the gravitation shifting, and also from the relativistic effects which are included by retaining ${\cal O}(v^{2}/c^{2})$ terms.}

We assume that resonant interaction of photons and matter takes place while propagating
only in the forward direction (in the direction towards the observer), and that all other possible interactions are neglected.
Altogether, these assumptions are known as the disconnected approximation \citep{GrachevGrinin,Marti77,RybickiHummer78,RybickiHummer83}. It is usually adopted in the multiple resonance case (see Paper I for the discussion of approximations). 

{\mybf Our computational domain has a spherical shape of a radius $p_{\rm out}$.
A ray with impact parameter, $p$ intersects this sphere in two places: $z_{1}$, and $z_{2}>z_{1}$. 
}
To obtain the emergent intensity, we sum the expression (\ref{Intens_jump_onEFS}) over all resonances along the ray with impact parameter $p$. 
{\mybf The total number of resonances is not known a priori and thus 
the procedure of the calculation of the total intensity is best cast in the form of a simple recurrence formula:
}
\begin{equation}\label{IntensityInfinity}
I_{i}(\nu^{\infty}, p)= {I}_{i-1} e^{-\tau_{{\rm l},i}}+
{S_{i}}(1-e^{-\tau_{{\rm l},i}})\mbox{,}
\end{equation}
where $i=1..N$, and $N$ is the total number of resonances, the initial intensity $I_{1}$ reads:

\begin{equation}
I_{1}= \left\{ I_{\rm c},\,p<R_{\rm c} \atop S_{0}  (1- e^{-\tau_{{\rm l}, 0}}  )\right.\mbox{,}
\end{equation}
where the subscript 0 refers to the point which is the farthest point along the ray from the observer in the computational domain 
{i.e. closest to $z_{1}$. Notice, that (\ref{IntensityInfinity}) is the summation formulae, it accounts for the fact that we don't know the total number of resonances before
we transfer a photon through the computational domain.
Thus we propagate the photon along it's trajectory (i.e. along a straight line with impact parameter $p$, and checking for the resonance with the line.
At the $i^{th}$ resonance we calculate $I_{i}$ , and go further; if another resonance is detected,
this, previous $I_{i}$ becomes a $I_{i-1}$ in a new iteration. Integration along the photon's trajectory 
is performed until the point $z_{2}$ is reached. All other couplings between resonances are neglected. Non-local coupling between resonances in the case of a normal hot-star, non-monotonic wind 
was investigated in \cite{Puls93}. From their results we may expect our approximation of the photon transfer is accurate within $10\%$.}

After being normalized to the continuum flux, $F_{\rm c}$  the radiation flux seen by the observer is calculated from the following:

\begin{equation}\label{flux_as_int_atinf}
F( \nu^\infty, p )/ F_{\rm c}= 
\int_0^{\infty}\,I^{\infty} (\nu^\infty, p)\,p\,dp
\mbox{,}
\end{equation}
where $I^{\infty}=I_{N}(\nu^{\infty}, p)$.

\section{Distributions of the velocity and opacity}
\begin{figure*}
\vspace*{174pt}
\includegraphics[width=450pt]{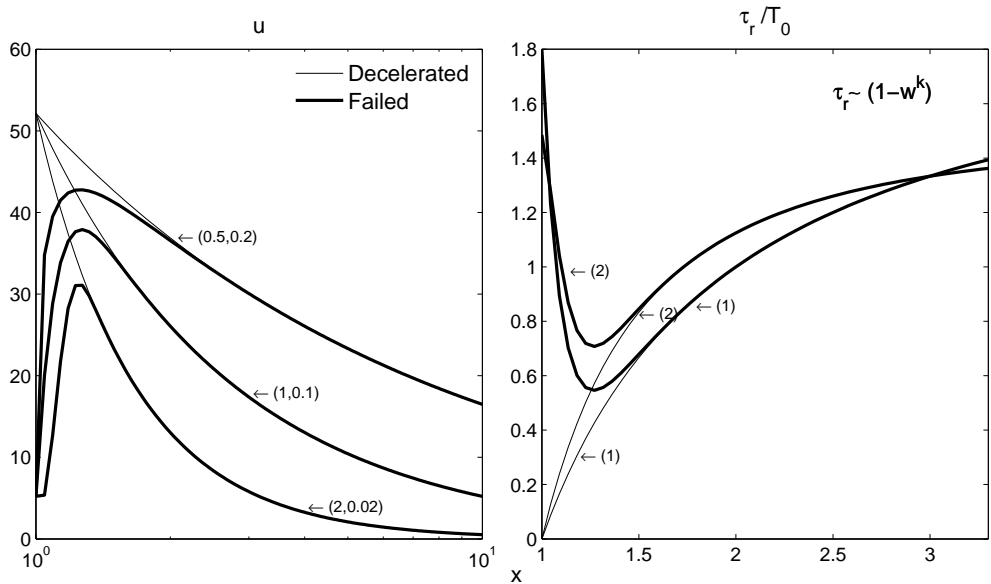}
\caption{
Velocity (left) and opacity (right) laws. 
Velocity curves: thin solid line: equation  (\ref{veloc_law1}); thick solid line: equation  (\ref{veloc_law2}); curves are labeled by
pairs of $m$, $\epsilon$.
Opacity curves calculated from the equation (\ref{tau_law1}) for corresponding velocity profiles;  thin solid line: using equation (\ref{veloc_law1}), and thick solid line: using equation (\ref{veloc_law2});
curves are marked by pairs of parameters $p$, and $\epsilon$ from equation (\ref{tau_law1})
}
\label{Figurevelocity_opacity1}
\end{figure*}

In this paper we consider two types of the velocity laws. Both are used to approximate a wind
which failed to escape from the potential well. The first adopted law describes the decelerating wind,

\begin{equation}\label{veloc_law1}
u(x)=U^{\infty}\frac{1}{x^{m}}\mbox{,}
\end{equation}
where $m>0$, and $U^{\infty}=V^{\infty}/v_{\rm th}$, $V^{\infty}$ is a terminal velocity. Equation (\ref{veloc_law1}) describes the wind, decelerating from $r=R_{\rm c}$ where its velocity has a maximum. A different approximation
is used to describe the wind which is accelerating at the beginning but after reaching maximum velocity is decelerating:

\begin{equation}\label{veloc_law2}
u(x)=U_{0}\left(\frac{1}{x^{m}}- (1-w_{0}/U_{0})e^{-(x-1)^m/\epsilon}  \right) \mbox{}
\end{equation}
where $w_0=v(R_{\rm c})/V^{\infty}=0.01$ and $0<\epsilon<0.1$ is related to the location or ''thickness'' of the boundary layer, which allows a smooth matching of the inner, increasing solution with the outer, decreasing one. 
The value of $U_{0}$ is related to $U^{\infty}$, and is calculated numerically so that $\max(u(x))=U^{\infty}$. 
In the acceleration part, in the limit $\delta x=x-1\sim 0$, equation (\ref{veloc_law2}) reduces to 
$u(x)\sim {\rm const}\cdot (\delta x)^{m} + w_{0}$,  and at infinity, $1/x^{m}$ term in the equation (\ref{veloc_law2}) is the dominant one. Velocity profiles (\ref{veloc_law1}) and (\ref{veloc_law2}) are shown in Figure \ref{Figurevelocity_opacity1} (left panel) for different values of $(m,\epsilon)$.
{\mybf In the Paper I, explosion-type outbursts (described by the homologous, Hubble-type velocity law), and winds with a stellar-type velocity profile 
were considered. With simplifications already adopted in this paper one can argue that there is no need in any further 
specification of the particular physical mechanism which could have lead to velocity laws (\ref{veloc_law1}), or 
(\ref{veloc_law2})  (although see discussion in Paper I). Having said that we consider them as a mere parameterization which may be relevant to the real failed or decelerated winds.}

The distribution of the opacity is adopted from \cite{CastorLamers}, who made use of the fact that in the optically thin case and the photoionization by diluted stellar continuum, the ionization ratio of the successive stages is proportional to $W/n$, where $W$ is the dilution factor and $n$ is the number density.
With the help of the continuity equation, in a spherically symmetric geometry, this ratio is proportional to $v$.  
It results in the parameterization of the radial optical depth, $\taur=\tau_{\rm l}(\mu=1)$ in terms of the non-dimensional velocity, $w\equiv v/V^{\infty}$, and is written in the following form: 
$\tau_{\rm r}\sim w$. {\mybf \citet{CastorLamers} also argued that $\tau_{\rm r}\sim 1-w$ law is somewhat more appropriate as 
the resultant P-Cygni profiles are closer to the observed ones. 
Notice that in the absence of self-consistent modeling of the opacity distribution together with the dynamics of the flow any such profiling of the opacity should be considered with caution. Most important for our means is weather $\tau_{\rm r}$ peaks close to the photosphere or at large radii.}.
{We chose the following parameterization of $\tau_{\rm r}$:}

\begin{equation}\label{tau_law1}
\taur(w)=T_0\frac{k+1}{k} (1-w^k)  \mbox{,}
\end{equation}

\noindent
where the parameter, $k\geq 0 $, and the parameter $T_0$ is related to the total optical depth at the line center,
$\displaystyle \int_{0}^{1}\,\tau_{\rm r}\,dw = T_{0}$.
Figure \ref{Figurevelocity_opacity1} (right panel) shows distributions of the opacity (\ref{tau_law1}) calculated from velocity laws (\ref{veloc_law1}), (\ref{veloc_law2}) for different pairs of $(m,\epsilon)$, and for $k=1$ and $k=2$.

\section{Surfaces of equal frequencies}
Equation (\ref{y_infty_main_equation}) determines the locus of equal frequency surfaces (EFS). That is, for a particular choice of the impact parameter,
$p$ and the frequency, $y^{\infty}$ this equation determines the position of the resonant point, $z_{0}$. In the Paper I it was shown, that in strong gravitational field, EFS have complicated shape and topology. 
Those of them which are calculated for the redshifted photons, may have branches situated between the star and observer. 
Of course, in the case of a normal stellar wind photons which are emitted in the gas which is approaching the observer, are blueshifted in the observer frame. 

In the language of resonant surfaces, it means there are no redshifted EFS in front of the star (we call redshifted or blueshifted EFS($\nu^{\infty}$) depending on whether $\nu^{\infty}$ is blue-, or redshifted). 
If gravitational redshifting
overwhelms Doppler boosting then in the observer frame these photons are redshifted. 

It is useful to  imagine resonant surfaces as partially transparent screens, which can scatter photons of a certain frequency. 
If EFS situates in front of the star, it produces an absorption line, according to $I_{0}\,e^{-\tau_{l}}$, where $I_{0}$ is the incident intensity and $\tau_{l}$ is the line optical depth,  
otherwise, EFS is reflecting, i.e.: $I=(1-e^{-\tau_{l}})\,S$, where $S$ is the source function. Notice that EFS also can be multi-branched. 

{\mybf
The topology and shape of resonant surfaces
in the case of the decelerated wind are different from those found in Paper I for the case of monotonically accelerated wind.
In Paper I it was found that sufficiently strong gravitational redshift can overwhelm doppler blueshift which results in a redshifted EFS positioned 
in front of the star.
}

\begin{figure*}
\vspace*{174pt}
\includegraphics[width=350pt]{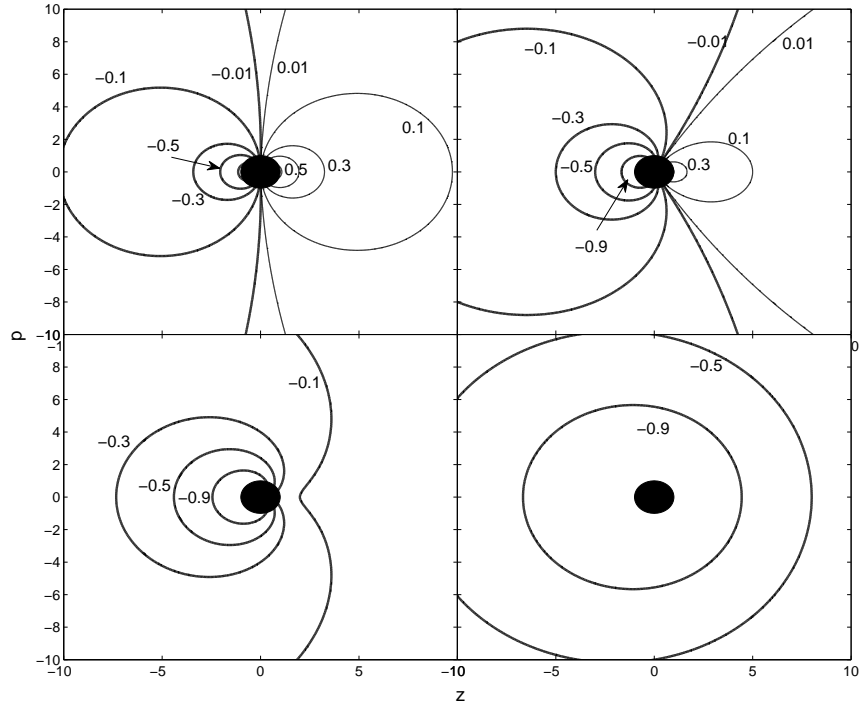}
\caption{
Equal frequency surfaces for the velocity law ~(\ref{veloc_law1}).
The parameters are: 
$g_{0}=2000$, $\beta_{\rm max}=0.01$ ({\it upper left}), 
$g_{0}=10$, $\beta_{\rm max}=0.1$ ({\it upper right}),
$g_{0}=8.33$, $\beta_{\rm max}=0.05$ ({\it lower left}),
$g_{0}=5$, $\beta_{\rm max}=0.02$ ({\it lower right}).
Curves are marked by the parameter $\Lambda$, corresponding to 
the non-dimensional frequency at infinity.
Curves: thick solid line: redshifted EFS; thin solid line: blueshifted EFS. 
The observer
is located at $z\to\infty$.}\label{FigureEFS_gam}
\end{figure*}

\begin{figure*}
\vspace*{174pt}
\includegraphics[width=350pt]{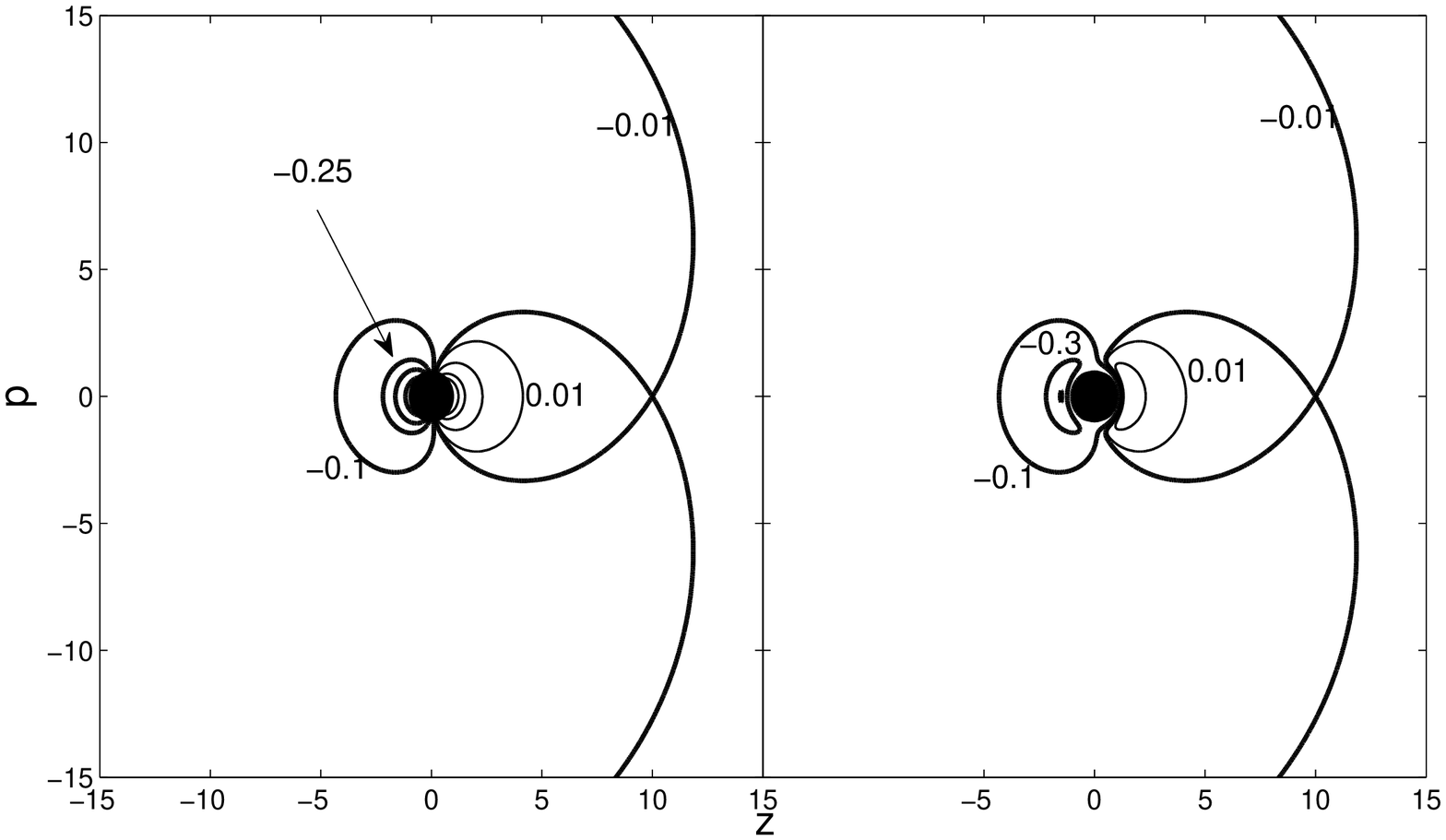}
\caption{Equal frequency surfaces; velocity law (\ref{veloc_law1}),({\it left}),
and (\ref{veloc_law2}), with $\epsilon= 0.1$ ({\it right}). Other parameters:  $m=2$, $g_{0}=16$, $\beta_{\rm max}=0.15$.
Thick solid line: redshifted EFS; thin solid line: blueshifted EFS. 
}\label{FigureEFS_1}
\end{figure*}

To calculate resonant surfaces 
we introduce the following non-dimensional parameters: 
$\Lambda=y^{\infty}/y^{\infty}_{\rm max}=y^{\infty}/u_{\rm max}$, where $u_{\rm max}=U^{\infty}$ in the case of 
the velocity law, (\ref{veloc_law1}), or the maximum velocity calculated from the velocity law, (\ref{veloc_law2}),
and the additional parameter $\beta_{\rm max}=(u_{\rm max}/c)v_{\rm th}$. Rewriting equation (\ref{y_infty_main_eq_nondim}),
we obtain:

\begin{equation}\label{y_infty_main_eq_nondim_2}
\Lambda = \mu w(x)+ \frac{1}{2(1-x g_{0})}\,\frac{1}{\beta_{\rm max}} -w(x)^{2}\beta_{\rm max}\mbox{.}
\end{equation}
{Equation (\ref{y_infty_main_eq_nondim_2}) is equivalent to equation (\ref{nu_infty_vs_nu_com}).
}
Notice that if no relativistic corrections are taken into account (i.e., no last term on the right), and additionally, the Newtonian potential is assumed,
the second term on the right becomes $-1/(2g_0 \, \beta_{\rm max})\, x^{-1}={\rm const}\times x^{-1}$, and the problem, is described in terms of only two parameters. 

Equal frequency surfaces calculated for the velocity law (\ref{veloc_law1}), are shown in Figure \ref{FigureEFS_gam}. Upper left plot shows EFS from the decelerated wind from the normal star, and its shape is well known, e.g. \citep{KuanKuhi, Marti77}. 

Increasing gravitational redshift, significantly changes the shape and position of EFSs, advancing the redshifted EFS in front of the star, 
and leading to formation of the redshifted absorption line or absorption edge.
 
Figure \ref{FigureEFS_1} shows EFS for velocity laws (\ref{veloc_law1}) and (\ref{veloc_law2}) for $g_{0}=16$, $m=2$. There are branches of both redshifted, and blueshifted EFS in front of the star, which makes possible the formation of both red- and blueshifted absorption lines (W-shape profiles). 
From Figure \ref{FigureEFS_1}, we see, that EFSs of the failed wind are not very different from those of the decelerated wind.

{\mybf Each node in the $\{p_{i}\}$ grid marks a ray, and
the problem of calculating the line profiles in $p,z$ geometry is reduced to summation of 
contributions from all such possible rays. Each such ray is discretized, $\{z_{k}\}$, and we iterate from $z_{1}$
to $z_{2}$ looking for resonances, and solving if necessary the non-linear equation (\ref{y_infty_main_eq_nondim}).
}

When integrating over the EFS one encounters a specific numerical problem related to sudden jumps of the brightness occurring 
at such $p_{i}(\nu^{\infty})$, at which the observer sees the boundary of the corresponding EFS($\nu^{\infty}$) 
(c.f. Figure \ref{FigureEFS_gam}, \ref{FigureEFS_1}, and also Paper I for discussion). 
If not treated properly, there are
spurious oscillations of the line profile observed for certain sets of parameters.
To some extent it may be cured by using finer discretization of the $\{p_{i}\}$ grid (see e.g. \cite{Marti77}). We prefer a different approach:
for each $\nu^{\infty}$ we numerically find such $(z_{j}, p_{j})$ at which $\partial p/\partial z=0$, and then, split the integral (\ref{flux_as_int_atinf}), into 
{two parts:
\begin{equation}\label{split_integral}
\int_{0}^{p_{j}}\,I^{\infty} (\nu^\infty, p)\,p\,dp+\int_{p_{j}}^{p_{\rm out}}\,I^{\infty} (\nu^\infty, p)\,p\,dp\mbox{,}
\end{equation}
which also allows to have two $\{p_{i}\}$ subgrids each having different mesh density. The latter is important for those cases in which different branches of EFS
are tightly packed near compact object. Thus we insure that $p_{j}$ is always exactly on the boundary between cells and thus, we eliminate spurious oscillations of the line profile.}

\section{Line profiles}
For the given set of parameters, $g_{0}$, $U^{\infty}$, $m$, $p$, and $\epsilon$
we calculate line profiles from equation (\ref{flux_as_int_atinf}) adopting velocity laws (\ref{veloc_law1}), (\ref{veloc_law2}), and the opacity 
law (\ref{tau_law1}). All profiles are calculated assuming the size of the computational domain,
$R_{\rm out} = 50\, R_{\rm c}$. In case of the failed wind law, (\ref{veloc_law1}),  the value of initial velocity is fixed constant, $w_{0}=0.01$, and in all cases $v_{\rm th} = 57\, {\rm km\,s^{-1}}$.

\subsection{Distorted P-Cygni and W-shaped profiles}

\begin{figure*}
\vspace*{174pt}
\includegraphics[width=450pt]{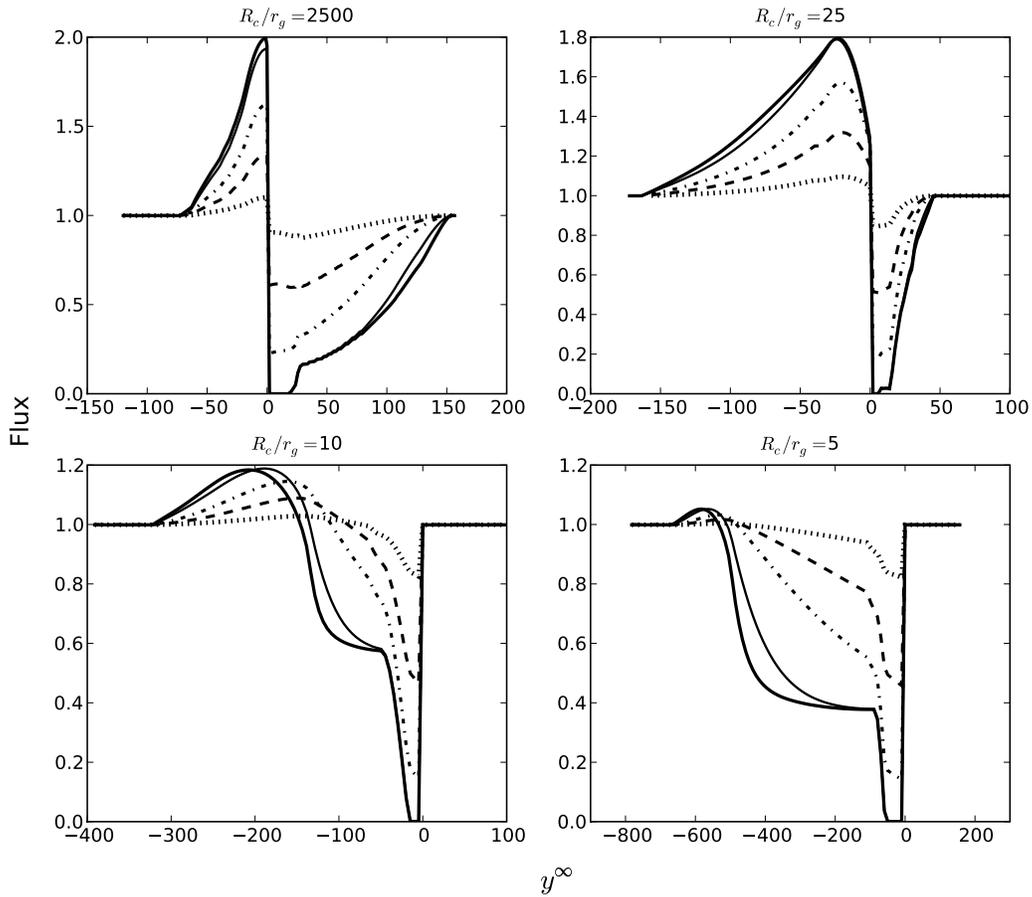}
\caption{Line profiles for different launching radii. Optical depth law: (\ref{tau_law1}), with $k=1$; velocity law:
(\ref{veloc_law1}), with $m=1$. Curves: dotted: $T_{0}=0.1$, dashed: $T_{0}=0.4$, 
dot-dashed: $T_{0}=1$; solid: thin: $T_{0}=4$; thick: $T_{0}=10$. Other parameters: $U^{\infty}=156.4$ corresponding to $V^{\infty}=0.03\,c$. 
}
\label{FigureLineProf_1}
\end{figure*}

\begin{figure*}
\vspace*{174pt}
\includegraphics[width=450pt]{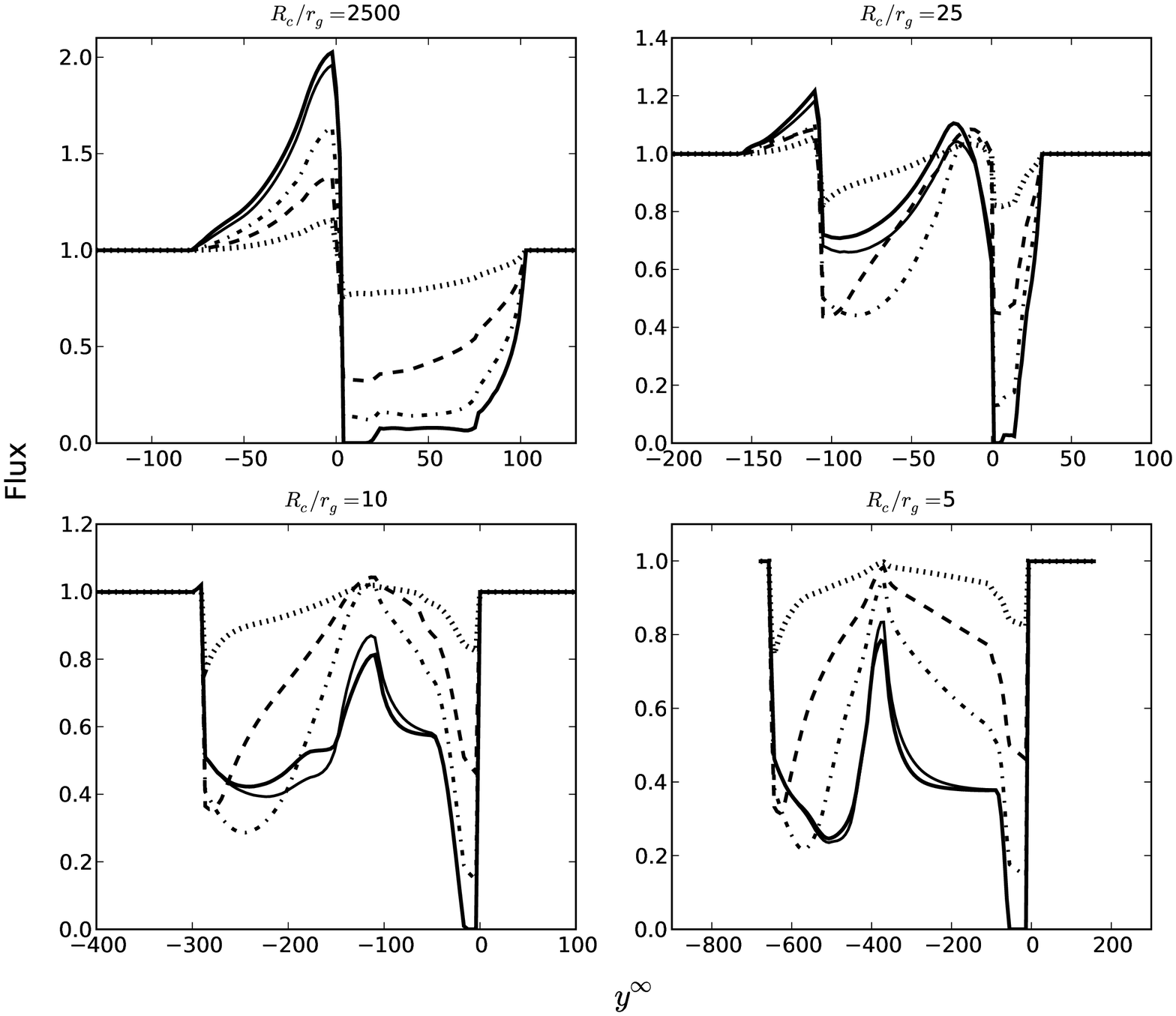}
\caption{Line profiles from failed wind approximated by (\ref{veloc_law2}), with $m=1$ and
$\epsilon=0.1$; $U_{0}=156.4$, which corresponds to $u_{\rm max}v_{\rm th}/c=0.02\,c$,  $u_{\rm max}=112$
(see text after (\ref{veloc_law2}) about the difference between  $U^{\infty}$ and $u_{\rm max}$, and $U_{0}$). 
Other parameters and the notation are the same as in Figure \ref{FigureLineProf_1}.}
\label{FigureLineProf_2}
\end{figure*}

Figures \ref{FigureLineProf_1} and Figure \ref{FigureLineProf_2} show profiles for different $g_{0}$, and velocity laws
(\ref{veloc_law1}), and (\ref{veloc_law2}).
From their corresponding upper left panels, we infer that both decelerating and failed winds show P-Cygni profile provided the influence of the gravitational redshifting is negligible. 

Notice that if equation (\ref{veloc_law2}) is adopted, the parameter $U_{0}$ does not equal the maximum velocity of the wind (see discussion after equation (\ref{veloc_law2})).
For example, Figures \ref{FigureLineProf_1} and \ref{FigureLineProf_2} are calculated for $U^{\infty}= U_{0}=156.4$, 
corresponding to $V^{\infty}=0.03 c$ in the case of Figure \ref{FigureLineProf_1}, and  $u_{\rm max}=112$, i.e. $u_{\rm max} \,v_{\rm th}/c=0.02\,c$ in the case of 
Figure \ref{FigureLineProf_2}.  
At large $g_{0}$ the maximum blueshift, i.e. $y^{+}_{\rm max}=y^{\infty}=V^{\infty}=156.4$, and 
the maximum projected velocity of the portion of the wind which is not obscured by the core determines the maximum observed redshifting: $y^{-}_{\rm max}\simeq -67.6$. 
Hereafter, $y^{\pm}$ is defined as a red- or blueshifted non-dimensional frequency $y^{\infty}$.
For the decelerating wind $u_{max}$ is reached very close to the bottom of the wind, where gravity is strong, opposite to the case considered in Paper I, where maximum velocity is reached far from the star where gravitational redshifting is negligible. 

The position of the red-ward edge of the line is determined from the approximate relation: $y^{+}_{\rm max}\simeq u_{\rm max} -\zeta/ (2\,g_{0})$. 
For $g_{0}=25$ we have $y^{+}_{\rm max}=52$, and from upper right of Figure \ref{FigureLineProf_1}, we infer $y^{+}_{\rm max}\simeq 46$. 
Comparing to Figure \ref{FigureLineProf_1} we see that this approximate formula works well if $u_{\rm max}$ is located close to the star.

The velocity law (\ref{veloc_law1}) generates the opacity distribution which has a maximum far from the star, while failed wind, (\ref{veloc_law2}) 
leads to the opacity law, which peaks close to the photosphere (c.f. Figure \ref{Figurevelocity_opacity1}).
There are no W-shape profiles in Figure \ref{FigureLineProf_1}.
If opacity peaks close to the photosphere, as in Figure \ref{FigureLineProf_2}, then profiles with two absorption troughs separated by ''emission'' component
are observed.

From Figure \ref{FigureLineProf_2} we can assume that W-shaped profiles are more likely to be observed from the failed wind
(\ref{veloc_law2}) rather than from the case of the decelerated wind, (\ref{veloc_law1}).

\begin{figure*}
\vspace*{174pt}
\includegraphics[width=450pt]{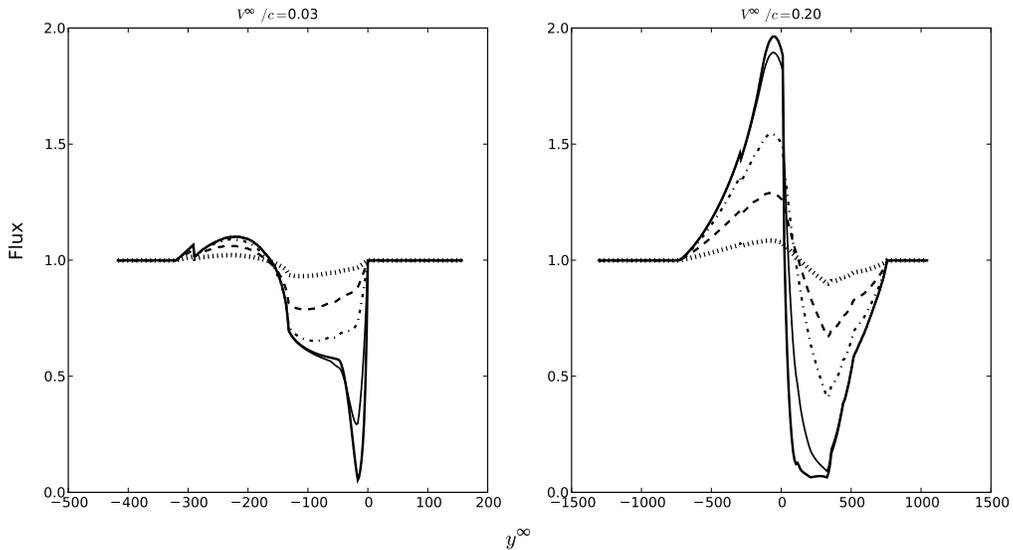}
\caption{Line profiles from decelerated wind with the velocity law (\ref{veloc_law1}), with $m=1$,
and opacity law (\ref{tau_law2});
both sets of profiles are calculated for $g_{0}=10$;
$V^{\infty} = 0.03\,c$ (left), $V^{\infty} = 0.2\,c$ (right). Curves: dotted: $T_{0}=0.1$, dashed: $T_{0}=0.4$, 
dot-dashed: $T_{0}=1$; solid: thin: $T_{0}=4$; thick: $T_{0}=10$.
}
\label{FigureLineProf_3}
\end{figure*}

To verify this assumption we calculate profiles for the velocity law (\ref{veloc_law1}) adopting opacity distribution which peaks at the photosphere in case of the velocity law (\ref{veloc_law1}):

\begin{equation}\label{tau_law2}
\taur(w)=T_0 (k+1) \,w^k  \mbox{,}
\end{equation}
The results are shown in Figure \ref{FigureLineProf_3}, where
both sets of profiles are calculated for $g_{0}=10$.

We conclude that a short acceleration phase of the failed wind is critical for the formation of the W-shaped profiles.

One can notice small spikes of intensity apparent in Figure  \ref{FigureLineProf_3}, they are most prominent in the absorption edge within the redshifted part of the profile (Figure \ref{FigureLineProf_2}, upper right).
In Figure  \ref{FigureLineProf_3}
at the redshifted part of the spectrum they are observed for both of $V^{\infty} = 0.03\,c$ (left), and 
$V^{\infty} = 0.2\,c$ (right). The position of such a spike is independent of $\tau$ and can be understood adopting the following considerations:
If gravitational shifting is negligible, a typical wind with $v\sim 1/r$ law has EFS which 
looks as two symmetrical loops in $p-z$ plane (for fixed $\pm \,y^{\infty}$) 
with respect of the $z=0$ plane (i.e. as in Figure \ref{FigureEFS_gam},  upper left). 

As $g_{0}$ decreases, the redshifted loop covers the star from the observer, and at higher redshifts (lower $g_{0}$) the loop shifts further behind the star, and covers only a fraction of it (Figure \ref{FigureEFS_1},  lower left).
This is due to the combination of the gravitational redshifting and 
strong Doppler red-, blueshifting at the photosphere. 
For lower $g_{0}$ (higher redshifts) 
EFS is obscuring less of a stellar disk, and such partial coverage results in interception of a lesser fraction of the direct flux of the core.  Additional flux in the spike comes from the continuum radiation scattered by the same EFS.

The position of such spikes, $y^{-}$ is approximately determined by gravitational and transverse Doppler redshift 
at the photosphere, $y^{-}\sim -\zeta/(2 g_{0}) - \beta^{2} \zeta/2$, at $x=1$; adopting parameters used to plot
Figure \ref{FigureLineProf_3} (left) we calculate $y^{-}\simeq -262$, while the figure gives $y^{-}=-291$. 

The velocity at $x\simeq1$ is high and even a small deviation of $\mu$ from zero, produces significant normal Doppler blue- or redshifting; the effect is larger at larger $v$, e.g. for the case of $V^{\infty} =0.2\,c$ (\ref{FigureLineProf_3}, right) we have: 
the predicted: $y^{-} \simeq -364$; 
the observed: $y^{-}\simeq -286$, which is closer to the shift produced by the gravitational field alone, $y^{-}_{\rm grav} \simeq -260$. From the above considerations it is clear that position of such spikes is independent of the actual distribution of opacity.

\begin{figure*}
\vspace*{174pt}
\includegraphics[width=450pt]{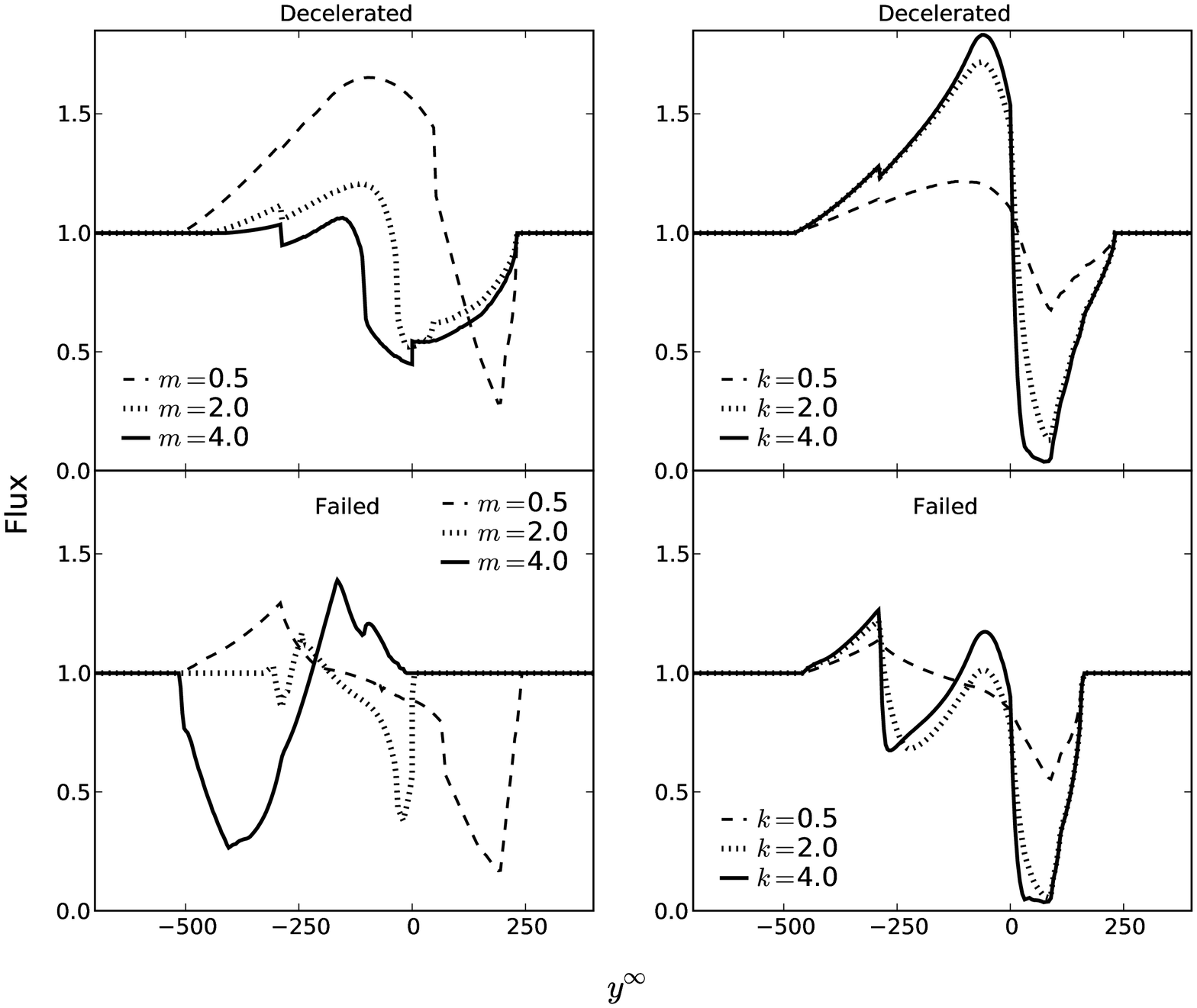}
\caption{Line profiles for different velocity laws. Upper panels:
curves for different $m$ (i.e. slope of the velocity profile) in (\ref{veloc_law1}), and (\ref{veloc_law2}), and 
fixed $k=1$ in (\ref{tau_law1}). Bottom panels: different $k$ (i.e. power index in  the opacity law) in (\ref{tau_law1}),
and fixed $m=1$ in (\ref{veloc_law1}) and (\ref{veloc_law2}).
Other parameters: $g_{0}=10$, $V^{\infty}=0.1\,c$, $T_{0}=1$, and $\epsilon=0.1$; if the velocity law (\ref{veloc_law2})
is adopted.
}
\label{FigureLineProf_4}
\end{figure*}

\subsection{Inverted P-Cygni profiles}
Figure \ref{FigureLineProf_4} shows profiles from failed and decelerated wind. 
In the case of the decelerated wind, 
varying $m$ in (\ref{veloc_law1}) or $k$ in (\ref{tau_law1}) results in the distortion of the P-Cygni profile. 
If the wind decelerates more 
rapidly (larger $m$) then the system of EFS is located closer to the star, and the lower surface area of EFS results in lower intensity of the profile 
(upper left). 
The effect from increasing $k$ in the opacity distribution is understood from Figure \ref{Figurevelocity_opacity1}, i.e. in the case of the velocity law (\ref{veloc_law1}) there is more opacity at the same radius, $x$ if $k$ is larger, resulting in that both the local contribution, $S(1-e^{-\tau_{\rm l}})$,
and the attenuation, $I_{\rm c}e^{-\tau_{\rm l}}$ is larger.

For the same setup of parameters shapes of profiles from the failed wind are completely different. 
From Figure \ref{FigureLineProf_4} (upper right) we see that as $m$ increases, the P-Cygni profile first transforms to the W-shaped, and then to the
{\it inverted} P-Cygni. 

To understand such behavior, we notice the W-shaped profiles are observed (at sufficiently strong gravity)
only in special cases of the accelerated wind (see Paper I), being very sensitive to the distribution of the opacity. 

In the case of accelerated wind in most cases there is a redshifted EFS behind the star at large $x$, because only Doppler redshifting is important  there. 
In result the surface area of the corresponding redshifted EFS is larger and its emission overwhelms the redshifted absorption. In the case of the decelerated or failed wind the maximum Doppler red-,  blueshifting is made by the gas which is close to the star. 

To produce an absorption line the EFS should be in front of the star and needs to have surface area to cover enough of a star. 
Roughly the same shape, low surface area EFS is situated behind 
of the star, and i) if it has small surface area it reflects only a relatively small fraction of the flux towards the observer, and ii) because of the obscuration by the core the flux received by the observer is much lower.

Thus, sufficiently strong gravity ''quenches'' the blueshifted EFS in front of the star, and thus there is no blueshifted absorption but only a strongly redshifted one. On both sides of the star Doppler blueshift from the gas is comparable with gravitational redshift (e.g. roughly at $\mu\lesssim\pi/2$).
This emission adds to the
radiation of the core. 
In result, the emission is blueshifted with respect to the redshifted absorption trough as shown in Figure \ref{FigureLineProf_4} (upper right).
Understanding that in the decelerated wind case the gravitational redshifting is capable of producing inverted P-Cygni profiles is one of the major findings of this work.

\begin{figure*}
\vspace*{174pt}
\includegraphics[width=450pt]{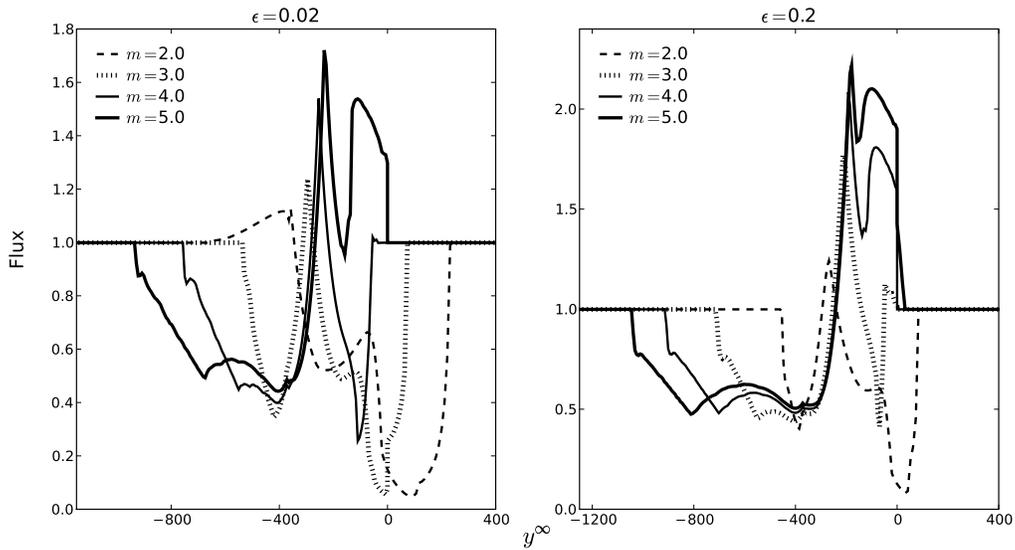}
\caption{Line profiles from failed wind (\ref{tau_law2}): left: $\epsilon=0.02$; right: $\epsilon=0.2$. 
Curves are for different values of $m$ at (\ref{tau_law2}).
Other parameters: $g_{0}=8$, $V^{\infty}=0.2\,c$. Opacity law (\ref{tau_law1}) with $k=1$.
}
\label{FigureLineProf_5}
\end{figure*}

It is instructive to investigate inverted P-Cygni profiles by changing the duration of the acceleration phase (i.e. $\epsilon$ in 
(\ref{veloc_law2})):
the smaller $\epsilon$, the shorter is such phase (Figure \ref{Figurevelocity_opacity1}).
In the case of the 
wind which is ''almost'' purely decelerating, $\epsilon=0.02$ the results are shown in Figure \ref{FigureLineProf_5} (left). 

Increasing $m$ has the following consequences: the distortion of P-Cygni profile (dashed); 
the formation of W-profile (dotted, thin solid), 
and finally the transformation to the inverted P-Cygni profile. The latter is characterized by a relatively weak absorption line superimposed on the blueshifted emission (thick solid).
In the case of  $\epsilon=0.2$ the behavior is qualitatively the same: modified P-Cygni, W-shaped, and inverted P-Cygni.

\section{Discussion}
Studying the influence of the gravitational redshifting on the shape of spectral line profiles
we adopt two basic scenarios for the wind dynamics. 
In the Paper I we calculated profiles from monotonically accelerated winds  (stellar-type winds, and homologous expansion).
It was established that gravitational redshifting is of significant importance, being able to modify and distorting line profiles considerably. 

Profiles are sensitive not only to wind dynamics but also to the distribution of the opacity.
Given the gravitational redshift is strong enough, and the terminal velocity, $V^{\infty}$ is either much smaller than the escape velocity at the base of the wind, or the distribution of the opacity strongly peaks close to the star, then a new type of profile, a W-shaped profile is produced.

We can imagine a situation (X-ray bursters as an example) when energy is deposited into the wind during a short initial pulse, and
after that time the wind is decelerating. In this paper we consider winds which are either decelerating, according to  $v(r)\sim 1/r^{m}$ law, or ''failed'', i.e. described by the bridging formula (\ref{veloc_law2}). In the latter case after a short accelerating phase, 
when $v(r)\sim ({r-R_{\rm c}})^{m}$, the wind velocity decreases according to $1/r^{m}$ law (c.f. Figure \ref{Figurevelocity_opacity1}). 
In both cases, the wind reaches its maximum velocity $V^{\infty}$ in the region where gravity is strong. 

{\mybf If there would be no absorption then the emission from the wind would be observed as a redshifted emission line having an approximately symmetric profile.
In the case of the stellar-type wind, $v(r)\sim(1-R_{\rm c}/r)^{\alpha}$ far from the compact object Doppler blueshifting always overwhelms gravity, and as a result there is always a possibility to have blueshifted absorption line. This absorption eats away blue-ward part of the emission line and produces a P-Cygni-type profile. On the other hand, inner parts of the wind also have velocities which are not large enough for the Doppler blueshifting to overwhelm gravitational redshifting, 
and such absorption eats away the red-ward part of the emission line. The result is that 
in certain cases these reshifted and blueshifted lines are combined together being separated by the emission component producing a W-shape profile.}

The case of the decelerated wind is different. Such a wind has its maximum velocity from the very beginning.
The maximum Doppler blueshift is reached close to the photosphere, where sufficiently strong gravitational redshifting can overwhelm it.
If there is an absorption it will be redshifted, and
so it is basically the regime where the photospheric line eats away the red-ward part of the emission from the wind. In such a case the inverted P-Cygni profile is produced. 
Depending on how quickly the wind decelerates, there also may be an absorption occuring far form the star, i.e. eating away the blue-ward part of the line and producing a W-shape profile similar to the the case of the monotonically accelerated wind.

Tree factors make inverted profiles 
possible: i)strong gravity which competes with the Doppler effect ii) the maximum velocity of the wind is approached in the region where gravitational redshifting is strong enough to overwhelm it, and iii) at larger radii the wind quickly decelerates.

If the energy is deposited into the wind during a short initial pulse
and the line is indeed formed in the vicinity of the compact object then it is likely to be transient, and
profiles similar to those obtained in this paper should be considered as snapshots of different stages of expansion. 
The profile of a line from the early stage of expansion would be different 
from the profile of the same line at later times. 
Though we believe that obtained profiles are robust, the need of deducing physical conditions of the line forming plasma in a real object would question simple approximations adopted in this paper, and such problem should be treated using methods of the 
full 3D modeling (such as in \citet{DorodnitsynKallman09}).

\section{Conclusions}
Several conclusion can be drawn about the influence of strong gravitational redshifting on profiles of spectral lines from failed and decelerated winds:
\begin{description}
\item[i)]  At large $R_{\rm c}/r_{g}$ where gravitational redshifting is negligible,
decelerated winds (\ref{veloc_law1}) and failed winds (\ref{veloc_law2}) produce profiles with
shapes similar to each other and to P-Cygni profile.
Profiles from the failed wind have absorption troughs which are flatter than those from the decelerating wind.

\item[ii)] 
If $R_{\rm c}$ is smaller than several tens of $r_{g}$ the effect of the gravitational shifting is 
significant. The strength of the effect
depends on the duration of the acceleration phase (parameter $\epsilon$): profiles from decelerated wind ($\epsilon =0$) are ''equivalent'' to P-Cygni, i.e. they have absorption trough which is blueshifted with respect to the emission peak.
Changing the slope of the velocity profile or changing the opacity distribution influences mostly the net intensity within the emission/absorption components; 
if $\epsilon> 0$ (failed wind) profiles are W-shaped, sometimes with additional weak emission component at the red-ward part of the profile
(saw-toothed profile). 
Such profiles are obtained in many cases when $R_{\rm c}$ is less than approximately 20-30 $r_{g}$. The effect of the gravity does not depend on $V^{\infty}$ as sensitively as in in the case of monotonically accelerated wind.


\item[iv)]
If failed wind is decelerating quickly (e.g. the parameter $m \gtrsim \,3-4$ in (\ref{veloc_law2}), for $R_{\rm c}=10\,r_{g}$, $V^{\infty}=0.1\,c$) the inverted 
P-Cygni profile is observed. 
Increasing the rate of the wind deceleration
(i.e. $m$), first produces the distorted P-Cygni profile, then  changes it to W-shaped, and then, at higher $m$ produces the absorption-emission inverted P-Cygni profile.
\end{description}

\hbox{}
This research was supported by an appointment at the NASA Goddard Space Flight Center, administered by CRESST/UMD through a contract with NASA, and by grants from the NASA Astrophysics Theory Program 05-ATP05-18.

\section*{Acknowledgments}

\bsp

\label{lastpage}

\end{document}